\documentclass[comment, prd,twocolumn,nofootinbib, preprintnumbers, superscriptaddress]{revtex4-2}
\usepackage{amsmath,amssymb,bm,slashed,braket}
\usepackage{graphicx}
\usepackage{epstopdf}
\usepackage{float}
\usepackage{xcolor}
\usepackage{multirow}
\usepackage{dsfont}
\usepackage[colorlinks=true,
            linkcolor=blue,
            urlcolor=blue,
            citecolor=purple,          
            bookmarks=true,
            bookmarksnumbered=true,
            breaklinks=true,
            pdfpagemode=Fullscreen,
            pdfstartview=FitBH]{hyperref}
\usepackage[normalem]{ulem}
\allowdisplaybreaks[4]
\usepackage[capitalise]{cleveref}
\usepackage{orcidlink}
\usepackage{microtype}

\usepackage{physics}
\usepackage[title]{appendix} 
\usepackage{subcaption}
\usepackage{tikz}

\begin{document}
\title{Studying magnetic circular vortex dichroism effect for photoionization of Rydberg atoms with vortex photons}
\author{Pengcheng Zhao\,\orcidlink{0000-0001-9211-6016}}
\email{20240264@m.scnu.edu.cn}
\affiliation{State Key Laboratory of Nuclear Physics and
	Technology, Institute of Quantum Matter, South China Normal
	University, Guangzhou 510006, China}
\affiliation{Guangdong Basic Research Center of Excellence for
	Structure and Fundamental Interactions of Matter, Guangdong
	Provincial Key Laboratory of Nuclear Science, Guangzhou
	510006, China}

\begin{abstract}
Rydberg atoms, renowned for their exceptional quantum properties, hold significant importance in quantum physics. The photoionization of Rydberg atoms serves as a critical tool for probing their unique characteristics.
In this work, we investigate the photoionization dynamics of {hydrogen-like Rydberg alkali atoms} interacting with vortex photons—a class of structured light carrying intrinsic orbital angular momentum. This {process} gives rise to novel quantum phenomena distinct from conventional photoionization processes.
Our results reveal that vortex photons exhibit exceptional sensitivity to the magnetic moments of Rydberg atoms, positioning them as a powerful spectroscopic tool for investigating Rydberg magnetism.
It is also demonstrated that the initial photon energy must be carefully selected to observe significant experimental results.
Furthermore, the photoionization process displays strong angular momentum selectivity, preferentially favoring configurations where the photon total angular momentum and atomic magnetic moment are aligned. 
This pronounced asymmetry directly manifests the chiral nature of the vortex photon-Rydberg atom {collisions}.

\end{abstract}

\maketitle 

\section{Introduction}
Since vortex photons were first shown to possess intrinsic orbital angular momentum (OAM) by Allen et al. in 1992 \cite{allen1992orbital}, they have found diverse applications across multiple domains. For instance, they serve as optical tweezers for nanoparticle manipulation \cite{simpson1997mechanical,curtis2002dynamic,leach2006optically}, offer advantages in optical imaging compared to conventional methods \cite{swartzlander2001peering,swartzlander2008astronomical,furhapter2005spiral,furhapter2005spiral2}, enable high-capacity optical communication due to their large discrete degrees of freedom \cite{gibson2004free,tamburini2012encoding,wang2012terabit,xie2016experimental}, and provide ideal systems for studying quantum entanglement \cite{mair2001entanglement,vaziri2002experimental,fickler2012quantum,malik2016multi}, among other applications.

Optical activity represents another important field where vortex photons demonstrate significant value. The circular dichroism (CD) effect, which describes the different absorption or scattering of left- and right-handed circularly polarized light, can be generalized to vortex photons \cite{forbes2021orbital}. This leads to the so-called vortex dichroism (VD) effect, characterizing the distinct interaction of vortex photons with opposite OAM states with matter, as demonstrated both theoretically and experimentally \cite{andrews2004optical,afanasev2017circular,zambrana2014angular}. When spin-orbit coupling is considered, the circular vortex dichroism (CVD) effect has also attracted considerable attention \cite{power1974circular,ye2019probing,kerber2018orbital}.

Magnetic dichroism effects, which put materials in magnetic fields and is the magnetic-field-induced counterparts of conventional dichroism effects, provide powerful tools for investigating magnetic properties. 
{Magnetic circular dichroism (MCD) effects, which characterize the different absorption of left- and right-handed circularly polarized light for material in magnetic fields, is well established as powerful tool for magnetic property detection~\cite{mason2007magnetic,snyder2007practical,van2014x}. 
Magnetic vortex dichroism (MVD) effects, which probe the dichroic response of matter to vortex beams carrying opposite OAM, have been proven to exhibit significantly stronger signals than MCD~\cite{van2007prediction,sirenko2019terahertz,fanciulli2021electromagnetic}.
Based on the extension of the conventional dichroism effect to CVD when both spin and orbital angular momentum are considered, we propose the magnetic circular vortex dichroism (MCVD) effect for magnetic systems. This phenomenon has not been investigated to date.}

{The collision between polarized vortex photons and polarized Rydberg atoms provides an ideal platform for studying the MCVD effect.
On the one hand, polarized vortex photons have been successfully generated and utilized in atomic photo-excitation processes~\cite{Schmiegelow:2016NC}. Polarized vortex states can be precisely controlled, and the atomic targets can be localized with sub-nanometer precision using optical trapping techniques.
On the other hand, the production and localization of polarized Rydberg atoms have also been achieved~\cite{Mehaignerie:2025PRXQ}. These atoms can be confined in carefully designed optical tweezers, with their polarization direction defined by a weak static magnetic field.
Given these advancements, it is now feasible to design experiments to study semi-central collisions between vortex photons and single Rydberg atom target. }

In this work, we investigate the MCVD effect by comparing the response of polarized Rydberg atoms to vortex photon beams with opposite topological charges and helicities. Through parity analysis, this comparison is shown to be equivalent to reversing the atoms' magnetic quantum numbers while keeping the photon beams unchanged. We theoretically calculate the photoionization cross-sections of Rydberg atoms interacting with vortex photons and present MCVD results for various initial topological charges and helicities. 
Specifically, initial vortex photon beam with energy $100$ eV is selected according to experimental condition~\cite{Bahrdt2013PRL111,Hemsing2013naturephysics9,Liu:2020wqd}.
The observed dependence of MCVD on the magnetic quantum number demonstrates the sensitivity of vortex photons to the magnetic moment of Rydberg atoms.
Furthermore, asymptotic behavior of MCVD for vortex photon with large total angular momentum (TAM) shows angular momentum selectivity of the process.

The paper is organized as follows:

Section \ref{section two} presents the transition amplitude and cross-section calculations.

Section \ref{section three} provides MCVD results and analyzes the observed phenomena.

Section \ref{section four} concludes the study.

Throughout this paper, three-dimensional vectors are denoted in boldface.
$\hbar=c=1$ is used.

\section{Theoretical calculations}\label{section two}
{In this paper, the target is one single hydrogen-like Rydberg alkali atom.
We investigate photoionization of its valence electron and use hydrogen-like wave function for calculations.
As being discussed in the introduction section, semi-central collision with such single atom target is feasible nowadays.  We begin with calculation of initial plane wave photon case and then go to initial vortex photon case.}
\subsection{Vortex states and collision scenario}
Prior to undertaking the calculations, it is instructive to briefly introduce the vortex states and the collision scenario that will be examined.
\subsubsection{Vortex states}
A vortex state typically refers to a type of structured wave that carries intrinsic OAM. It features a singularity line where the amplitude vanishes and the phase becomes indeterminate. The probability current of such a state flows helically around this singularity line, providing a visual representation of its intrinsic OAM.

\begin{figure}[htbp]
    \centering
    \includegraphics[width=0.9\linewidth]{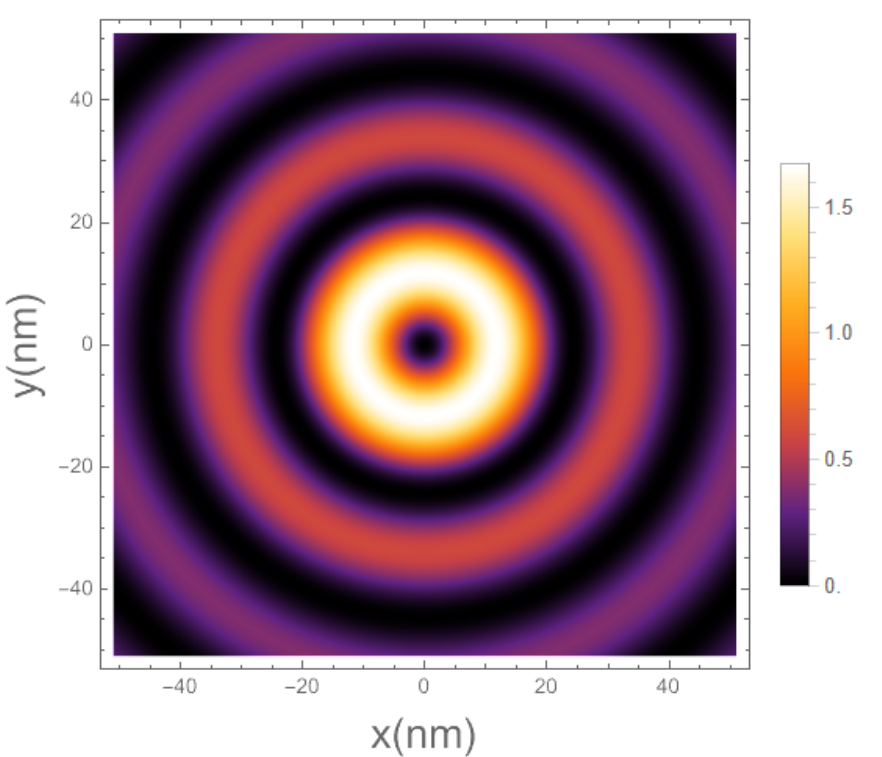}
    \caption{\raggedright Transverse intensity profile of a Bessel vortex photon with $\omega =100$ eV, $\theta_{\vb k}=\pi/10$, $m_{\gamma}=1$, and $\Lambda_{\gamma}=-1$. All transverse distances are in nanometers.}
    \label{Bessel profile}
\end{figure}

For our purposes, the most convenient type is the Bessel vortex state, which will be used throughout this paper. A Bessel vortex state for a photon is given by~\cite{Afanasev:2013PRA}
\begin{eqnarray}\label{Bessel}
    \vb A ^{B}(\vb r,\,t)&=&\int \frac{d^2\vec k_{\perp}}{(2\pi)^2}\sqrt{\frac{2\pi}{\kappa}}(-i)^{m_{\gamma}} {\vb {\epsilon}}_{\mathbf k{\Lambda_{\gamma}}}\delta (|\vec k_{\perp}|-\kappa){\rm e}^{i\,({\vb k} \cdot {\vb r}-\omega t)}\nonumber\\
    &=&N_B\int \frac{d\phi_{\vb k}}{2\pi}{\rm e}^{i\,m_{\gamma}\phi_{\vb k}}{\rm e}^{i\,\vb k \cdot \vb{r}}{\vb {\epsilon}}_{\vb k{\Lambda_{\gamma}}},
\end{eqnarray}
where $N_B=(-i)^{m_{\gamma}}\sqrt{{\kappa}/{2\pi}}$ is the normalization factor, and $\vec k_{\perp}$ represents the transverse part of the momentum $\mathbf k$ (its longitudinal part is denoted $k_z$). In spherical coordinates, the momentum is parameterized as $\vb{k} = (k, \theta_{\vb k}, \phi_{\vb k})$. The state is characterized by four quantum numbers: the energy $\omega$, the transverse momentum $\kappa$, the total angular momentum (TAM) projection $m_{\gamma}$, and the helicity ${\Lambda_{\gamma}}$. It carries a TAM of $\hbar m_{\gamma}$, where $m_{\gamma}$ can be any integer. The quantity $\bar m_{\gamma}=m_{\gamma}-\Lambda_{\gamma}$ is defined as the topological charge, which dictates the phase variation of the state. The transverse intensity profile consists of bright and dark concentric rings with a central null point. An example for a Bessel vortex state is shown in Fig.~\ref{Bessel profile}.

\begin{figure}[htbp]
    \centering
    \includegraphics[width=0.9\linewidth]{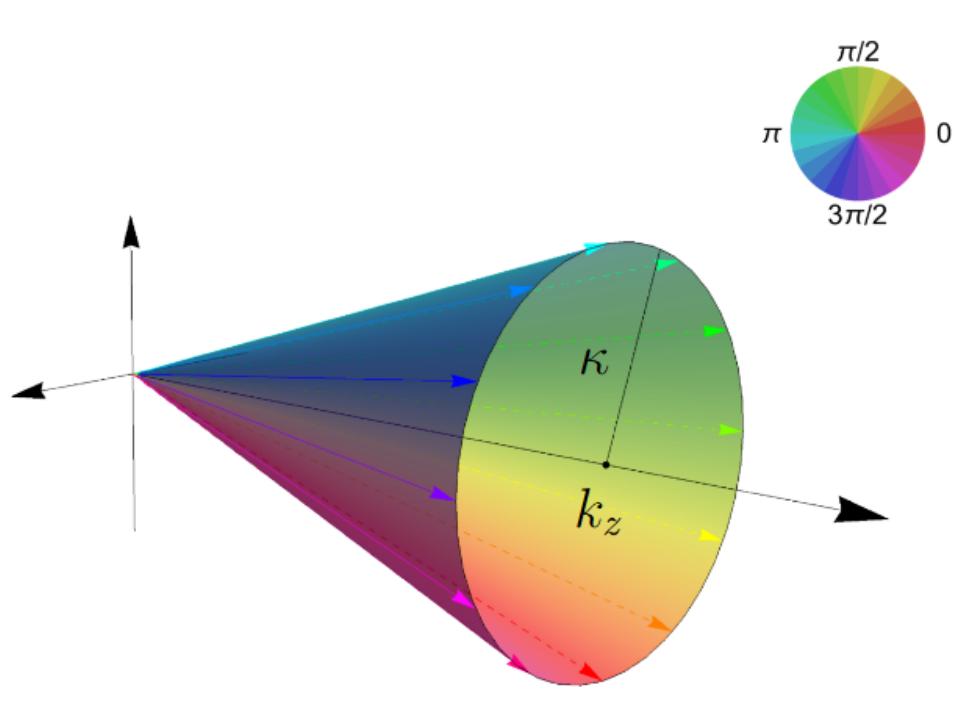}
    \caption{\raggedright Momentum distribution of a monochromatic Bessel vortex state with topological charge $\bar m_{\gamma}=1$. The black circle of radius $\kappa$ indicates the distribution in momentum space. All momentum vectors lie on a circular cone. The phase factor of the plane wave components advances by $2\pi \bar m_{\gamma}$ around the circle (phases are color-coded). Here, $k_z$ is the longitudinal momentum and $\kappa$ is the fixed transverse momentum magnitude for all components.}\label{bessel vortex state}
\end{figure}

This state is a superposition of multiple plane wave components. In momentum space, these components are distributed on a circle orthogonal to the propagation direction, and their momentum vectors collectively form a circular cone, as depicted in Fig. \ref{bessel vortex state}.

Finally, the position-dependent photon current density is given by \cite{knyazev2018beams}:
\begin{align}
    j^V(r_{\perp}) &= \frac{k\cos\theta_{\vb k}}{2\pi}\Bigg[ \cos^4\frac{\theta_{\vb k}}{2}J_{m_{\gamma}-{\Lambda_{\gamma}}}^2(k_{\perp}r_{\perp}) \nonumber\\
    &\quad+ \frac{1}{2}\sin^2\theta_{\vb k} J_{m_{\gamma}}^2(k_{\perp}r_{\perp}) + \sin^4\frac{\theta_{\vb k}}{2}J^2_{m_{\gamma}+{\Lambda_{\gamma}}}(k_{\perp}r_{\perp})\Bigg] ,
\end{align}
where $r_{\perp}$ denotes the transverse coordinate and $J_n(x)$ is the Bessel function of the first kind.
\subsubsection{Collision scenario}
\begin{figure}[htbp]
    \centering
    \includegraphics[width=0.9\linewidth]{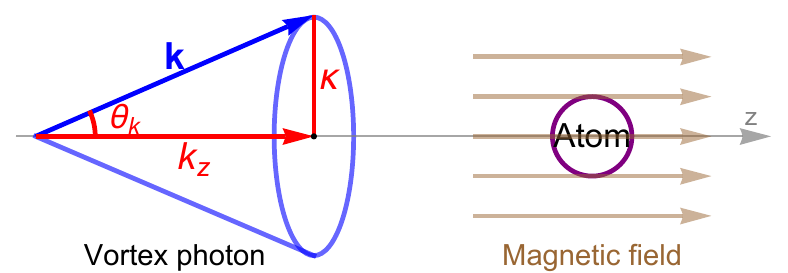}
    \caption{\raggedright Geometry of a central collision between a vortex photon and a Rydberg atom. The polarized Rydberg atom is located within a weak, constant magnetic field. The incident vortex photon (represented by the blue cone with cone angle $\theta_k$) propagates along an axis collinear with the magnetic field direction. Here, $k_z$ and $\kappa$ denote the photon's longitudinal momentum and transverse momentum modulus, respectively. The atom is positioned directly on the singularity line of the vortex photon.}
    \label{collision scenario}
\end{figure}

For a central collision between a vortex photon and a Rydberg atom, the scattering geometry is illustrated in Fig.~\ref{collision scenario}.
In this setup, the target atom is placed within a weak, constant magnetic field that defines its quantization axis. A vortex photon is incident on the atom, with the atom located precisely on the singularity line of the vortex beam. The direction of the magnetic field and the propagation direction of the vortex photon are collinear.

\subsection{Transition matrix for Rydberg photoionization with initial plane wave photon}
Under the first Born approximation, the single-electron photoionization matrix element for an atomic state by a plane wave photon in coordinate space is given by:

\begin{equation}\label{transition matrix coordinate space}
{\cal M}_{fi}^P(\vb{k},\vb{p}) = -\frac{e}{m_e}\int \dd[3]{\vb{r}}\,
\psi^*_f(\vb{r}) e^{i\vb{k} \cdot \vb{r}} 
(\vb{\epsilon}_{\vb{k}{\Lambda_{\gamma}}}\cdot \hat{\vb{p}})\psi_{nlm_l}(\vb{r}),
\end{equation}
{where $e$ is the electron charge, $m_e$ is the electron mass, $\vb{r}$ is the position coordinate, $\vb{k}$ is the incident photon momentum, $\vb{\epsilon}_{\vb{k}{\Lambda_{\gamma}}}$ is the photon polarization vector with helicity ${\Lambda_{\gamma}}$, $\vb{p}$ is the outgoing electron momentum, $\hat {\vb p}=-i \vb {\nabla}$ is the momentum operator, $\psi_f$ is the final electron state and $\psi_{nlm_l}$ is the initial bound state.}
{$(n,l,m_l)$ are principle quantum number, orbital quantum number and magnetic quantum number of atomic state respectively.}

In spherical coordinates, the momenta are parameterized as $\vb{k} = (k, \theta_{\vb k}, \phi_{\vb k})$ and $\vb{p} = (p, \theta_{\vb p}, \phi_{\vb p})$. {The photon polarization vector in Coulomb gauge takes the explicit form~\cite{Afanasev:2013PRA}:}

\begin{equation}
\vb{\epsilon}_{\vb{k}{\Lambda_{\gamma}}} = \begin{pmatrix}
-\dfrac{{\Lambda_{\gamma}}}{\sqrt{2}}e^{-i{\Lambda_{\gamma}}\phi_{\vb k}}\cos^2\dfrac{\theta_{\vb k}}{2} 
+ \dfrac{{\Lambda_{\gamma}}}{\sqrt{2}}e^{i{\Lambda_{\gamma}}\phi_{\vb k}}\sin^2\dfrac{\theta_{\vb k}}{2} \\[10pt]
-\dfrac{i}{\sqrt{2}}e^{-i{\Lambda_{\gamma}}\phi_{\vb k}}\cos^2\dfrac{\theta_{\vb k}}{2} 
- \dfrac{i}{\sqrt{2}}e^{i{\Lambda_{\gamma}}\phi_{\vb k}}\sin^2\dfrac{\theta_{\vb k}}{2} \\[10pt]
\dfrac{{\Lambda_{\gamma}}}{\sqrt{2}}\sin\theta_{\vb k}
\end{pmatrix},
\end{equation}
{where $\Lambda _{\gamma} $ is helicity of the state.}

Choosing the final state as a plane wave $\psi_f^* = e^{-i\vb{p}\cdot\vb{r}}$, we obtain the momentum-space representation:

\begin{equation}\label{transition amplitude in momentum space}
{{\cal M}_{fi}^P(\vb k,\vb p) = -\frac{e}{m_e}(\vb{\epsilon}_{\vb{k}{\Lambda_{\gamma}}}\cdot \vb{q})
\tilde{\psi}_i^{nlm}(\vb{q}),}
\end{equation}
where $\vb{q} \equiv \vb{p} - \vb{k}$ is the momentum transfer and $\tilde{\psi}_i$ denotes the Fourier transform of the initial state.

The Coulomb gauge condition $\vb{\epsilon}_{\vb{k}{\Lambda_{\gamma}}}\cdot\vb{k} = 0$ implies:
\begin{align}\label{epsilon p}
\vb{\epsilon}_{\vb{k}{\Lambda_{\gamma}}}\cdot \vb{q} &= \vb{\epsilon}_{\vb{k}{\Lambda_{\gamma}}}\cdot \vb{p} \nonumber \\
&= \frac{{\Lambda_{\gamma}}}{\sqrt{2}}p \Bigg[ \sin\theta_{\vb p}\bigg(\sin^2\frac{\theta_{\vb k}}{2}e^{i{\Lambda_{\gamma}}(\phi_{\vb k}-\phi_{\vb p})} \nonumber \\
&\quad - \cos^2\frac{\theta_{\vb k}}{2}e^{-i{\Lambda_{\gamma}}(\phi_{\vb k}-\phi_{\vb p})}\bigg) 
+ \sin\theta_{\vb k}\cos\theta_{\vb p} \Bigg],
\end{align}
where $p \equiv \norm{\vb{p}}$ is the electron momentum magnitude.

{We investigate the photoionization of the valence electron in alkali metal atoms, where only the outermost electron is ionized. Due to their hydrogen-like character, the ionization process can be described using hydrogen atom wave functions.}
{
The hydrogen-like atomic wavefunction in coordinate space is
\begin{align}
    {\psi}_{nlm_l}(\vb r)&=\sqrt{\Big( \frac{2}{na} \Big) ^3\frac{(n-l-1)!}{2n(n+l)!}}{\rm e}^{-r/na}\Big( \frac{2r}{na}\Big) ^l\nonumber\\
    &\quad \times \Big[ L_{n-l-1}^{2l+1}\Big(\frac{2r}{na}\Big)\Big] Y_l^{m_l}(\theta,\phi),
\end{align}
where $a$ is the Bohr radius, $(r,\theta,\phi)$ are spherical coordinates, $L_{\alpha}^{\beta}$ is the associated Laguerre polynomials and $Y_{l}^{m_l}$ is the spherical harmonic function. The atom's polarization axis has been chosen to be $z$ axis.}
By Fourier transform, we get its formula in momentum space\cite{Matula2013Atomic}:
\begin{align}
\tilde{\psi}_{nlm_l}(\vb{q}) &= 2^{2l+2} l! (-i)^l n^{-l-2}
\sqrt{\frac{2(n-l-1)!}{\pi (n+l)!}} 
\frac{1}{(q^2a^2 + \delta^2)^{l+2}} \nonumber \\
&\quad \times C_{n-l-1}^{l+1}\left( \frac{q^2a^2 - \delta^2}{q^2a^2 + \delta^2} \right) 
( {qa} )^l Y_{l}^{m_l}(\theta_{\vb q}, \phi_{\vb q}),
\end{align}
{where $\delta=1/n$ and $C_n^\alpha(x)$ is the Gegenbauer polynomial.  }

\subsection{Transition matrix for Rydberg photoionization with initial vortex photon}

\subsubsection{Momentum Space Wavefunction for the selected Rydberg atoms}
For simplicity, we choose $n=l+1$.
The atomic wavefunction in momentum space is given by:
\begin{align}\label{atom}
\tilde{\psi}_{n{{(n-1)}}{m_l}}(\vb{q}) &= 2^{2n} l! (-i)^l n^{-l-2}\sqrt{\frac{2}{\pi (2l+1)!}} \nonumber\\
& \quad \times \frac{q^l a^l Y_{l}^{m_l}(\theta_{\vb q}, \phi_{\vb q})}{(q^2a^2 + \delta^2)^{l+2}} \nonumber \\
&= \hat{N}\frac{q^l Y_{l}^{m_l}(\theta_{\vb q}, \phi_{\vb q})}{(n^2q^2a^2 + 1)^{l+2}},
\end{align}
where the normalization constant is:
\begin{equation}
\hat{N} = 2^{2n} l! (-i)^l n^{l+2}a^l\sqrt{\frac{2}{\pi (2l+1)!}}.
\end{equation}

\subsubsection{Plane Wave Photoionization for the selected Rydberg atoms}
{According to Eq.~\ref{transition amplitude in momentum space}, Eq.~\ref{epsilon p} and Eq.~\ref{atom}}, the transition amplitude for plane wave photons is:
\begin{align}\label{mp}
{\cal M}_{fi}^P(\vb{k},\vb{p}) &= -\frac{\hat{N} e{\Lambda_{\gamma}}}{\sqrt{2}m_e} \Bigg[ p_{\perp}\sin^2\frac{\theta_{\vb k}}{2}{\rm e}^{i{\Lambda_{\gamma}}\phi}- p_{\perp}\cos^2\frac{\theta_{\vb k}}{2}{\rm e}^{-i{\Lambda_{\gamma}}\phi} \nonumber \\
&\quad + p_z\sin\theta_{\vb k} \Bigg] \frac{q^l Y_{l}^{m_l}(\theta_{\vb q}, \phi_{\vb q})}{(n^2a^2q^2 + 1)^{l+2}},
\end{align}
where $\phi=\phi_{\vb k}-\phi_{\vb p}$, $p_{\perp}=p\sin \theta_{\vb p}$ and $p_z=p\cos \theta_{\vb p}$.
Specially, when photon momentum is parallel with polarization axis of the target atom (i.e. $\theta_{\vb k}=0$), we have
\begin{align}\label{mp0}
{\cal M}_{fi}^P(\vb{k},\vb{p}) &= \frac{\hat{N} e{\Lambda_{\gamma}}}{\sqrt{2}m_e}   p_{\perp}\frac{q^l Y_{l}^{m_l}(\theta_{\vb q}, \phi_{\vb q})}{(n^2a^2q^2 + 1)^{l+2}}.
\end{align}
This equation shows that $|{\cal M}_{fi}^P|$ is independent of photon helicity ${\Lambda_{\gamma}}$ under the first Born approximation.

For simplicity, we write Eq.~\ref{mp} as:
\[{\cal M}_{fi}^P(\vb{k},\vb{p}) =N_P\mathcal{A}(\theta_{\vb k},\phi)\frac{q^l Y_{l}^{m_l}(\theta_{\vb q}, \phi_{\vb q})}{(n^2a^2q^2 + 1)^{l+2}},\]
where
\[N_P=-\frac{\hat{N} e{\Lambda_{\gamma}}}{\sqrt{2}m_e} \]
and 
\[\mathcal{A}(\theta_{\vb k},\phi) =p_{\perp}\sin^2\frac{\theta_{\vb k}}{2}{\rm e}^{i{\Lambda_{\gamma}}\phi}- p_{\perp}\cos^2\frac{\theta_{\vb k}}{2}{\rm e}^{-i{\Lambda_{\gamma}}\phi} + p_z\sin\theta_{\vb k}.\]

\subsubsection{Vortex Photon Photoionization for the selected Rydberg atoms}
For central collision between vortex photon and Rydberg atom, the amplitude is obtained through azimuthal averaging:
\begin{align}\label{mv}
{\cal M}^V_{fi}(\vb p) &= N_B\int_0^{2\pi} \frac{d\phi_{\vb k}}{2\pi} {\rm e}^{im_{\gamma}\phi_{\vb k}} M^P_{fi}(\vb{k}, \vb{p})\nonumber\\
&=N_BN_P{\rm e}^{im_{\gamma}\phi_{\vb p}}\int_0^{2\pi} \frac{d\phi}{2\pi} {\rm e}^{im_{\gamma}\phi}\mathcal{A}(\theta_{\vb k},\phi)\nonumber\\
&\quad \times \frac{q^l Y_{l}^{m_l}(\theta_{\vb q}, \phi_{\vb q})}{(n^2a^2q^2 + 1)^{l+2}}.
\end{align}

Using the solid harmonic expansion:
\begin{equation}
q^lY_{l}^{m_l}(\theta_{\vb q},\phi_{\vb q}) = \sqrt{\frac{2l+1}{4\pi}}R_{l}^{m_l}(\vb{q}),
\end{equation}
and the addition theorem for solid harmonics:
\begin{align}
R_{l}^{m_l}(\vb{r}+\vb{a})& = \sum_{{{\rho}}=0}^l\sum_{\mu=-{{\rho}}}^{{{\rho}}} {\sqrt{\binom{2l}{2{{\rho}}}}} R_{{{\rho}}}^{\mu}(\vb{r})R_{l-{{\rho}}}^{{m_l}-\mu}(\vb{a})
\nonumber\\
&\quad \times \braket{{{\rho}},\mu;l-{{\rho}},{m_l}-\mu}{l{m_l}}
\end{align}
where ${\binom{2l}{2{{\rho}}}}$ is the binomial coefficient,
we obtain the expanded formula:
\begin{align}\label{qY}
\frac{q^l Y_{l}^{m_l}(\theta_{\vb q}, \phi_{\vb q})}{(n^2a^2q^2 + 1)^{l+2}}&= \frac{{\rm e}^{i{m_l}\phi_{\vb p}}}{(n^2a^2q^2+1)^{l+2}} \nonumber \\
&\quad \times \sum_{{{\rho}}=0}^l\sum_{\mu=\mu _{\rm min}}^{\mu _{\rm max}} f_{l{m_l},{{\rho}}\mu}(\vb{p}) e^{i({m_l}-\mu)\phi},
\end{align}
where the coefficient function is:
\begin{align}
f_{l{m_l},{{\rho}}\mu}(\vb{p}) &= \sqrt{\frac{(2l+1)({{\rho}}-\mu)!(l-{{\rho}}-{m_l}+\mu)!}{4\pi({{\rho}}+\mu)!(l-{{\rho}}+{m_l}-\mu)!}\binom{2l}{2{{\rho}}}} \nonumber \\
&\quad \times p^{{{\rho}}}(-k)^{l-{{\rho}}} P_{{{\rho}}}^{\mu}(\cos\theta_{\vb p})P_{l-{{\rho}}}^{{m_l}-\mu}(\cos\theta_{\vb k}) \nonumber \\
&\quad \times \braket{{{\rho}},\mu;l-{{\rho}},{m_l}-\mu}{l{m_l}}.
\end{align}
$P_{{{\rho}}}^{\mu} $ is the associated Laguerre polynomial and the limitation of $\mu$ is: $\mu_{\rm min}={\rm Max}(-l+{m_l}+{{\rho}},\,-{{\rho}})$ and $\mu_{\rm max}={\rm Min}(l+{m_l}-{{\rho}},\,{{\rho}})$.

After substituting Eq.\ref{qY} into Eq.\ref{mv} and performing the variable change $z = e^{i\phi_{\vb{k}}}$, we obtain:
\begin{align}\label{mvfi}
{\cal M}_{fi}^V(\vb p) &= \frac{i(-1)^n N_BN_P}{2\pi(na)^{2l+4}} e^{i({m_l}+m_{\gamma })\phi_{\vb{p}}} \sum_{{{\rho}}=0}^l\sum_{\mu=\mu _{\rm min}}^{\mu _{\rm max}}  f_{l{m_l},{{\rho}}\mu}(\vb{p}) \nonumber \\
&\quad \times \oint_{\mathcal{C}} dz \, \frac{\mathcal{B}(z)}{(p_{\perp}k_{\perp})^{l+2} z^{-\zeta}(z-z_1)^{l+2}(z-z_2)^{l+2}},
\end{align}
where $\zeta ={m_l}+m_{\gamma }+l-\mu$ and 
\[\mathcal{B}(z) = p_{\perp}\sin^2(\theta_{\vb{k}}/2) z^{1+{\Lambda_{\gamma}}} - p_{\perp}\cos^2(\theta_{\vb{k}}/2)z^{1-{\Lambda_{\gamma}}} + p_z\sin\theta_{\vb{k}} \, z,\] 
and $\mathcal{C}$ denotes the unit circle in the complex plane.

The contour integral in Eq.~\eqref{mvfi} can be evaluated using the residue theorem. The explicit analytical result is presented in Appendix~\ref{appendix a}.

\subsection{Differential Cross Section}
The photoionization differential cross section is given by
\begin{equation}\label{cross_section}
    d\sigma = \frac{|{\cal M}|^2}{j}(2\pi)\delta(E_f - E_i - \omega)\frac{d^3\vb{p}}{(2\pi)^3},
\end{equation}
where $E_f$ is the kinetic energy of the final electron, $E_i$ is the binding energy of the initial electron, $\omega$ is the energy of the incident photon, and $j$ represents the incident photon current. 
For different initial photon state, current $j$ will be different.
{Here, $\cal M$ can be ${\cal M}^P_{fi}(\vb k,\vb p)$ in Eq.~\ref{transition matrix coordinate space} or ${\cal M}_{fi}^V(\vb p)$ in Eq.~\ref{mv}.}

When $|{\cal M}|$ is independent of the azimuthal angle $\phi_{\vb p}$ of the outgoing electron, we obtain the angular differential cross section:
\begin{equation}
    \frac{d\sigma}{d\theta_{\vb p}} = \frac{ m_e p \sin\theta_{\vb p}}{2\pi j}|{\cal M}|^2.
\end{equation}

Here, for initial vortex photon, we use an effective photon current:
\[ j^V_{\rm eff} = \int j^V(r_{\perp})|\psi_i|^2(\vb {r})d^3\vb{r}. \]
The efficiency of this current is discussed in Appendix.\ref{appendix b}.

\subsection{{{{{{{{Multipole}}}}}} expansion and approximations}}
{In the next section, we will demonstrate a range of phenomena using the exact calculations described above. Prior to that, however, it will be useful to introduce an approximation method based on {{{{{{multipole}}}}}} expansion, which will aid in analyzing the upcoming phenomena.}

\subsubsection{{{{{{{multipole}}}}}} expansion}
We can perform a {{{{{{multipole}}}}}} expansion for the photon field (see formula $(16.58)$ in reference \cite{Thephysicsofatomsandquanta}):
\begin{align}
{{\rm e}^{i\vb k \cdot \vb r}=\sum _{{{\eta}} =0}^ {\infty}\frac{1}{{{\eta}} !}(i\vb k \cdot \vb r)^{{{\eta}}}.}
\end{align}
Then Eq.~(\ref{transition matrix coordinate space}) becomes
\begin{align}
	\mathcal{M}_{fi}^P(\vb k,\vb p) &= -N_B\frac{e}{m_e}\sum_{{{\eta}}=0}^{\infty} \Bigg[ \frac{1}{{{\eta}}!} \int d^3\vb{r}\, e^{-i\vb{p}\cdot\vb{r}} (i\vb{k}\cdot\vb{r})^{{{\eta}}} \nonumber\\
    &\quad \times (\vb{\epsilon}_{k{\Lambda_{\gamma}}} \cdot \hat{\vb{p}}) \psi_i(\vb{r}) \Bigg] \nonumber \\
	&= -N_B\frac{e}{m_e}\sum_{{{\eta}}=0}^{\infty} \left[ \frac{1}{{{\eta}}!} (\vb{\epsilon}_{k{\Lambda_{\gamma}}} \cdot \vb{p}) (-\vb{k} \cdot \vb{\nabla}_{\vb p})^{{{\eta}}} \tilde{\psi}_i(\vb{p}) \right],
\end{align}
where 
\begin{align}\label{multiple}
	\vb k\cdot \vb {\nabla}_{\vb p} &= k(\sin{\theta_{\vb k}}\sin{\theta_{\vb p}}\cos{\phi} + \cos{\theta_{\vb k}}\cos{\theta_{\vb p}})\partial_{ p} \nonumber \\
	&\quad + \frac{k}{p}(\sin{\theta_{\vb k}}\cos{\theta_{\vb p}}\cos{\phi} - \cos{\theta_{\vb k}}\sin{\theta_{\vb p}})\partial_{\theta_{\vb p}} \nonumber \\  
	&\quad + \frac{k}{p}\frac{\sin{\theta_{\vb k}}}{\sin{\theta_{\vb p}}}\sin{\phi}\partial_{\phi_{\vb p}},
\end{align}
with $\phi = \phi_{\vb k} - \phi_{\vb p}$.
In particular, the azimuthal dependence enters only in the form $\phi_k-\phi_p$ due to rotational invariance.

The transition matrix for the vortex case is
\begin{align}\label{vortex}
	\mathcal{M}^V_{fi}(\vb p) &= N_B\int \frac{d\phi_{\vb k}}{2\pi} e^{i m_{\gamma}\phi_{\vb k}} \mathcal{M}^P_{fi}(\vb k) \nonumber \\
	&= -N_B\frac{e}{m_e}\sum_{{{\eta}}=0}^{\infty} \Bigg[ \frac{1}{{{\eta}}!} \int \frac{d\phi}{2\pi} e^{i m_{\gamma}\phi_{\vb k}} (\vb{\epsilon}_{k{\Lambda_{\gamma}}} \cdot \vb{p}) \nonumber\\
    &\quad \times(-\vb{k} \cdot \vb{\nabla}_p)^{{{\eta}}} \tilde{\psi}_i(\vb{p}) \Bigg] \nonumber \\
	&= -N_B\frac{e}{m_e} e^{i m_{\gamma}\phi_{\vb p}} \sum_{{{\eta}}=0}^{\infty} \Bigg[ \frac{1}{{{\eta}}!} \int \frac{d\phi}{2\pi} e^{i m_{\gamma}\phi} (\vb{\epsilon}_{k{\Lambda_{\gamma}}} \cdot \vb{p}) \nonumber\\
    &\quad \times (-\vb{k} \cdot \vb{\nabla}_{\vb p})^{{{\eta}}} \tilde{\psi}_i(\vb{p}) \Bigg].
\end{align}

By considering Eq.~(\ref{epsilon p}), (\ref{multiple}), and (\ref{vortex}), we can analyze the dependence of {{{{{{multipole}}}}}} contributions on TAM of the vortex photon. 
For the ${{\eta}}$-th order {{{{{{multipole}}}}}} contribution:
\begin{itemize}
	\item The factor $(\vb{\epsilon}_{\vb k{\Lambda_{\gamma}}} \cdot \vb{p})$ contains $\phi$-dependent phase terms $e^{i m_{\Lambda}\phi}$ with $m_\Lambda=0,~\pm 1$
	\item {The factor $(-\vb{k} \cdot \vb{\nabla}_p)^{{{\eta}}}$ contains $\phi$-dependent phase terms $e^{i m_{\nabla}\phi}$ with $m_\nabla=0,~\pm 1,~\cdots,~\pm{{\eta}}$}
\end{itemize}
The global $\phi$-dependent phase factor $e^{i(m_{\gamma} + m_{\Lambda} + m_{\nabla})\phi}$ serves as a filter when integrated over one complete cycle. This integration selects only terms satisfying the conservation condition: $m_{\gamma} + m_{\Lambda} + m_{\nabla} = 0$, which constrains the photon TAM to $m_{\gamma} = -m_{\Lambda} - m_{\nabla}$, with allowed values spanning $0$ to $\pm ({{\eta}} + 1)$. The {{{{{{multipole}}}}}} expansion exhibits the following characteristic hierarchy:
\begin{itemize}
    \item Dipole(${{\eta}} = 0$): $m_{\gamma} \in \{0, \pm 1\}$
    \item Quadrupole(${{\eta}} = 1$): $m_{\gamma} \in \{0, \pm 1, \pm 2\}$
    \item Hexapole(${{\eta}} = 2$): $m_{\gamma} \in \{0, \pm 1, \pm 2\, \pm 3\}$
\end{itemize}
This pattern extends to higher {{{{{{multipole}}}}}} orders.
For photo-ionization with vortex photon at fixed $m_{\gamma }$, the minimal contributing order is ${{\eta}} = |m_{\gamma}| - 1$ for $|m_{\gamma}| \geq 1$, while the $m_{\gamma} = 0$ case begins with ${{\eta}} = 0$.
\subsubsection{Approximations}
Since the operator $\partial_p$ always yields a $1/p$ term when acting on $\tilde{\psi}_{nlm_l}$, we can extract a global factor $k/p$ from Eq.~\ref{multiple}. This factor is typically small for non-relativistic photoionization processes. Consequently, higher-order terms generally has smaller contributions to Eq.~\ref{vortex}. 

With the $\phi$ integral, we explicitly get $m_{\nabla}=-m_{\gamma }-m_{\Lambda}$.
The lowest order contribution has ${{\eta}}=|m_{\nabla}|=|m_{\gamma }+m_{\Lambda}|$.
Eq.~\ref{multiple} contains $\phi$-independent terms that has no contribution to the lowest order term, due to the filtering effect of $\phi$ integral. We can therefore neglect these terms when evaluating Eq.~\ref{vortex}, leading to the simplified expression:
\begin{align}\label{multiple2}
    \vb k\cdot \vb {\nabla}_{\vb p} &\approx k\sin{\theta_{\vb k}}\sin{\theta_{\vb p}}\cos{\phi}\,\partial_{p} 
    + \frac{k}{p}\sin{\theta_{\vb k}}\cos{\theta_{\vb p}}\cos{\phi}\,\partial_{\theta_{\vb p}}\nonumber\\
    &\quad
    + \frac{k}{p}\frac{\sin{\theta_{\vb k}}}{\sin{\theta_{\vb p}}}\sin{\phi}\,\partial_{\phi_{\vb p}}.
\end{align}

For Rydberg atoms, the wave function in momentum space is given by Eq.~\ref{atom}:
\begin{align}
\tilde{\psi}_{nl{m_l}}(\vb{p}) \propto \frac{p^l Y_{l}^{m_l}(\theta_{\vb p}, \phi_{\vb p})}{(n^2p^2a^2 + 1)^{l+2}}.
\end{align}
For Rydberg atoms with large $n$, the term $n^2p^2a^2$ is much larger than unity, allowing the approximation:
\begin{align}\label{approx}
\tilde{\psi}_{nl{m_l}}(\vb{p}) \propto \frac{Y_{l}^{m_l}(\theta_{\vb p}, \phi_{\vb p})}{n^{2(l+2)}a^{2(l+2)}p^{l+4}}.
\end{align}
Consequently, the action of differential operators on $\tilde{\psi}_{nlm_l}$ takes specific forms: $\partial{p} \rightarrow -(l+4)/p$, $\partial_{\theta_{\vb p}} \rightarrow l\cot{\theta_{\vb p}}$, and $\partial_{\phi_{\vb p}} \rightarrow i{m_l} = i\lambda l$, where we have defined ${m_l} = \lambda l$ with $\lambda = \pm 1$ corresponding to positive and negative ${m_l}$ values, respectively.
Thus, we obtain exact formula for the {{{{{{multipole}}}}}} operators:
\begin{align}\label{Rydbergapprox}
    -\vb k\cdot \vb {\nabla}_{\vb p} &\approx \frac{k\sin\theta_{\vb k}}{p} \Big\{  \Big[ (l+2)\sin\theta_{\vb p}-\frac{l}{\sin \theta_{\vb p}}\Big] e^{i\lambda\phi} \nonumber\\
    &\quad +(l+2)\sin\theta_{\vb p}e^{-i\lambda\phi}\Big\}, \nonumber\\
    (-\vb k\cdot \vb {\nabla}_{\vb p})^2 &\approx \frac{k^2\sin^2\theta_{\vb k}}{p^2} \Big\{  \Big[ (l+2)\sin\theta_{\vb p}-\frac{l}{\sin \theta_{\vb p}}\Big] e^{i\lambda\phi} \nonumber\\
    &\quad +(l+2)\sin\theta_{\vb p}e^{-i\lambda\phi}\Big\} \Big\{  \Big[ (l+\frac{5}{2})\sin\theta_{\vb p}\nonumber\\
    &\quad -\frac{l}{\sin\theta_{\vb p}}\Big] e^{i\lambda\phi} 
     +(l+\frac{5}{2})\sin\theta_{\vb p}e^{-i\lambda\phi}\Big\}\nonumber\\
    (-\vb k\cdot \vb {\nabla}_{\vb p})^{{{\eta}}} &\approx \left(\frac{k\sin\theta_{\vb k}}{p}\right)^{{{\eta}}} \prod_{j=1}^{{{\eta}}} \Big\{  \Big[ (l+\frac{3+j}{2})\sin\theta_{\vb p}\nonumber\\
    &\quad -\frac{l}{\sin\theta_{\vb p}}\Big] e^{i\lambda\phi}  +(l+\frac{3+j}{2})\sin\theta_{\vb p}e^{-i\lambda\phi}\Big\}.
\end{align}
Specially for CRA, the electron wavefunction concentrates near $\theta_{\vb p}\approx \pi/2$, we obtain further approximations for the {{{{{{multipole}}}}}} operators:
\begin{align}\label{craapprox}
    -\vb k\cdot \vb {\nabla}_{\vb p} &\approx \frac{k\sin\theta_{\vb k}}{p} \left[ (l+2)e^{-i\lambda\phi} + 2e^{i\lambda\phi} \right], \nonumber\\
    (-\vb k\cdot \vb {\nabla}_{\vb p})^2 &\approx \left(\frac{k\sin\theta_{\vb k}}{p}\right)^2 \left[ (l+2)e^{-i\lambda\phi} + 2e^{i\lambda\phi} \right] \nonumber\\
    &\quad \times \left[ (l+2.5)e^{-i\lambda\phi} + 2.5e^{i\lambda\phi} \right], \nonumber\\
    (-\vb k\cdot \vb {\nabla}_{\vb p})^{{{\eta}}} &\approx \left(\frac{k\sin\theta_{\vb k}}{p}\right)^{{{\eta}}} \prod_{j=1}^{{{\eta}}} \Bigg[ \left(l+\frac{j+3}{2}\right)e^{-i\lambda\phi} \nonumber\\
    &\quad + \left(\frac{j+3}{2}\right)e^{i\lambda\phi} \Bigg].
\end{align}
The higher-order contributions are strongly suppressed by the $(k\sin\theta_{\vb k}/p)^{{\eta}}$ factor, with the $\sin\theta_{\vb k}$ term providing additional suppression.

\section{Results and Discussions}\label{section three}
\subsection{Magnetic circular-vortex dichroism effect}
The \textit{magnetic circular dichroism} (MCD) effect manifests distinct absorption or scattering probabilities for left- and right-circularly polarized photons incident on chiral targets in a static magnetic field. This phenomenon has emerged as a powerful tool for probing the magnetic properties of materials.
For vortex photons, the analogous \textit{magnetic vortex dichroism} (MVD) effect arises from different absorption or scattering of optical vortices carrying opposite OAM (or topological charge).
We introduce a novel dichroic phenomenon, termed \textit{magnetic circular-vortex dichroism} (MCVD), which characterizes the different response of a magnetized target to polarized vortex beams with opposite TAM and helicity:{
\begin{align}\label{mcvd definition}
{ \mathrm{MCVD}= \frac{\sigma(  m_{\gamma},\Lambda_{\gamma}=1) - \sigma(- m_{\gamma},\Lambda_{\gamma}=-1)}{\sigma( m_{\gamma},\Lambda_{\gamma}=1) + \sigma(-m_r,\Lambda_{\gamma}=-1)} }.
\end{align}}
{It is a physical observable measured via a two-collision process, in which vortex beams with opposite TAM and helicity are directed at the same polarized atom. And it is the core subject of this work.}
{For simplicity, in the following text, we use "$(m_{\gamma},\Lambda_{\gamma})$" to refer to the MCVD in different initial parameters defined in Eq.~\ref{mcvd definition}.}

{Note that we use the eigenstates quantized by four quantum numbers: energy, longitudinal momentum, TAM and helicity. OAM is not a good quantum number in general. Only for the paraxial case that OAM can approximately be a good quantum number. The two states involving in MCVD has opposite TAM and helicity. Their OAM is determined by relation "TAM=OAM+helicity" for paraxial case.}

\paragraph*{Symmetry analysis}
In the case of Rydberg atom photoionization by vortex photons, MCVD can be studied by inverting the photon's quantum numbers---the TAM quantum number ${m}_{\gamma}$ and helicity ${\Lambda_{\gamma}}$---while keeping the target state fixed. 
This configuration is equivalent to comparing ionization from two target states, $\ket{\psi_{n,l,{m_l}}}$ and $\ket{\psi_{n,l,-{m_l}}}$, under identical photon conditions.

The underlying symmetry corresponds to a mirror inversion in the transverse plane, equivalent to a one-dimensional parity transformation. 
Under this operation, factors in Eq.~\ref{mv}:
\begin{itemize}
    \item The phase factors ${\rm e}^{i\phi}/{\rm e}^{i\phi_{\vb{q}}}$ transform as ${\rm e}^{-i\phi}/{\rm e}^{-i\phi_{\vb{q}}}$ (up to a constant phase)
    \item The length $q$ remains unchanged
    \item The differential $d\phi$ changes sign ($d\phi \to -d\phi$)
\end{itemize}

These transformations preserve the modulus of the matrix element $\mathcal{M}^V_{fi}$. 
Physically, this operation reverses all angular momentum-related quantum numbers: ${m}_{\gamma}$, ${\Lambda_{\gamma}}$, and ${m_l}$. 
For the specific case of $(-{m}_{\gamma}, -{\Lambda_{\gamma}}, {m_l})$, the transition probability $\abs{\mathcal{M}^V_{fi}}^2$ becomes identical to that of $({m}_{\gamma}, {\Lambda_{\gamma}}, -{m_l})$.
MCVD is therefore given by:{
\begin{align}\label{mcvd}
\mathrm{MCVD}&= \frac{\sigma( m_{\gamma},\Lambda_{\gamma}=1,{m_l}) - \sigma(- m_{\gamma},\Lambda_{\gamma}=-1,{m_l})}{\sigma( m_{\gamma},\Lambda_{\gamma}=1,{m_l}) + \sigma( -m_{\gamma},\Lambda_{\gamma}=-1,{m_l})} \nonumber \\
&=\frac{\sigma( m_{\gamma},\Lambda_{\gamma}=1,{m_l}) - \sigma( m_{\gamma},\Lambda_{\gamma}=1,-{m_l})}{\sigma( m_{\gamma},\Lambda_{\gamma}=1,{m_l}) + \sigma( m_{\gamma},\Lambda_{\gamma}=1,-{m_l})}.
\end{align}
}
This demonstrates that MCVD measurement requires only target inversion along the photon's propagation direction for experimental verification.

\subsection{$m_l$ dependence of MCVD for Rydberg Atoms}

To investigate the sensitivity of vortex photons to the magnetic moment of Rydberg atoms, we analyze the dependence of MCVD on the magnetic quantum number ${m_l}$ under two distinct experimental configurations. In both cases, we consider several types of vortex photons characterized by different TAM quantum number $ m_{\gamma}$ and helicity ${\Lambda_{\gamma}}$.
{A Rydberg atom with main quantum number $n=51$ is investigated. Its radius is about $138$nm. Initial vortex photon has energy $\omega=100$eV and conical angle $\theta_k=\pi/10$. In these parameters, the transverse characteristic length for vortex photon is about $365$nm which is in the same order with Rydberg atom's radius.}

\begin{figure}[htbp]
    \centering
    \includegraphics[width=0.9\linewidth]{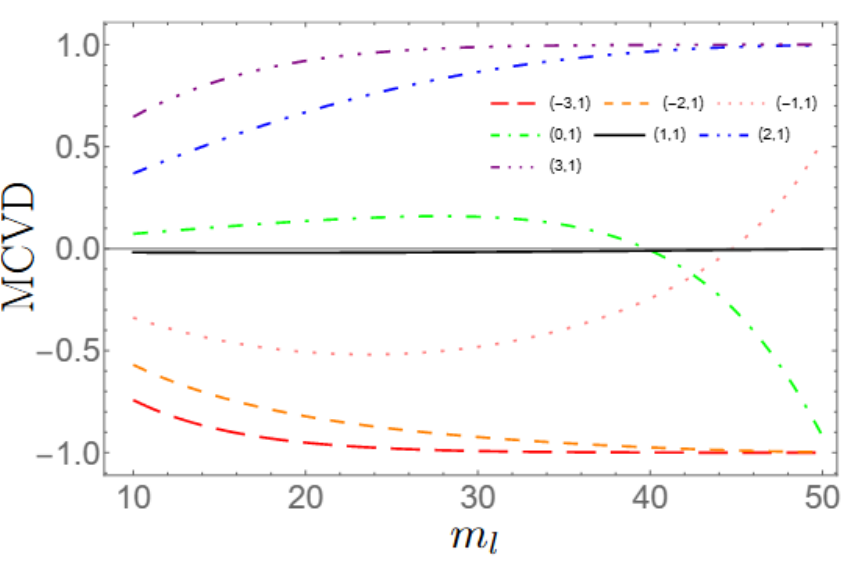}
    \caption{\raggedright MCVD in Rydberg atom photoionization with vortex photons. The principal and orbital quantum numbers are fixed at $n = l + 1 = 51$, while the magnetic quantum number varies from $10$ to $50$. Initial photon energy is chosen to be $100$ eV. Conical angle of vortex photon is $\theta_k=\pi/10$. Different lines corresponds to different cases with $m_{\gamma}$ ranging from $0$ to $\pm 3$. Label $(a,b)$ in the legends means we get MCVD from difference of two initial cases: $(m_{\gamma}=a,\Lambda_{\gamma}=b)$ and ($m_{\gamma}=-a,\Lambda_{\gamma}=-b)$.} 
    \label{fig:mcd_nl}
\end{figure}
\paragraph{First Configuration: Fixed Principal and Orbital Quantum Numbers}
In the first scenario, we fix the principal ($n$) and orbital ($l$) quantum numbers while varying the magnetic quantum number ${m_l}$. Figure~\ref{fig:mcd_nl} presents the MCVD curves for different initial photon states, considering Rydberg atoms with $n = l + 1 = 51$. For MCVD which we have defined in Eq.~\ref{mcvd}, we examine vortex photons with TAM quantum number ranging from $m_{\gamma}=0$ to $m_{\gamma}=\pm 3$. 

The MCVD approaches zero for the case $m_{\gamma }=1$, closely resembling the plane-wave photon scenario (Eq.~\ref{mp0}) where the transition amplitude modulus becomes helicity-independent at tree level and the MCVD is zero.  
Other cases exhibit significant MCVD sensitivity to ${m_l}$, directly reflecting the atomic magnetic moment. 
For $|m_{\gamma}| \geq 2$, the MCVD shows monotonic ${m_l}$-dependence, confirming its effectiveness for Rydberg atom magnetic moment measurements. At large ${m_l}$, MCVD asymptotically approaches 1, revealing enhanced chiral asymmetry for states with higher magnetic quantum numbers. The $|m_{\gamma}| = 2$ case demonstrates stronger MCVD variations than $|m_{\gamma}| = 3$, making it preferable for analysis. While cases with $m_{\gamma}=-1,\,0$ display non-monotonic trends, their high-$m_l$ tails show significant $m_l$ dependence.  

\begin{figure}[htbp]
    \centering
    \includegraphics[width=0.9\linewidth]{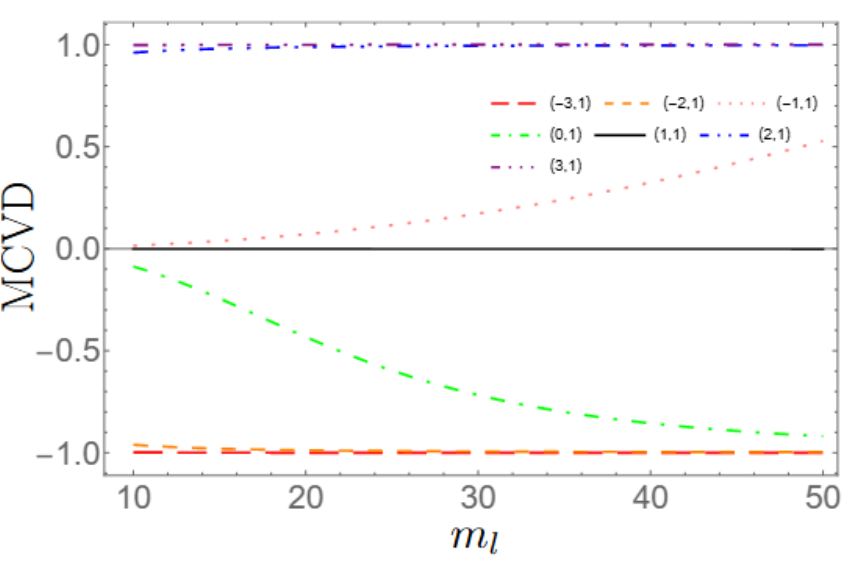}
    \caption{\raggedright MCVD in CRA photoionization with vortex photons. Targets follow $n=l+1={m_l}+1$ with $n$ varying from $11$ to $51$. Initial photon energy is chosen to be $100$ eV.  Conical angle of vortex photon is $\theta_k=\pi/10$.  Different lines corresponds to different cases with $m_{\gamma}$ ranging from $0$ to $\pm 3$. Label $(a,b)$ in the legends means we get MCVD from difference of two initial cases: $(m_{\gamma}=a,\Lambda_{\gamma}=b)$ and ($m_{\gamma}=-a,\Lambda_{\gamma}=-b)$.  }
    \label{fig:mcd_cra}
\end{figure}
\paragraph{Second Configuration: CRAs With Variable Principal Quantum Numbers}
{In the first configuration, the shape of the wave function varies with the magnetic quantum number $m_l$. In contrast, the second scenario employs CRAs with different principal quantum numbers $n$ to isolate the effect of $m_l$ under a unchanging wave function shape.} With the magnetic quantum number fixed as ${m_l}=n-1$ (considering only ${m_l}\geq 0$), this configuration effectively studies ${m_l}$-dependence.
Figure \ref{fig:mcd_cra} presents MCVD curves with ${m_l}$ ranging from $10$ to $50$.
Again, for MCVD which we have defined in Eq.~\ref{mcvd}, we examine vortex photons with TAM quantum number ranging from $0$ to $\pm 3$.  

The $m_{\gamma}=1$ case again shows MCVD approaching zero, demonstrating that vortex photons with TAM matching their helicity ($m_\gamma = {\Lambda_{\gamma}}$) behave similarly to plane-wave photons. 
The $m_{\gamma}=-1$ and $m_{\gamma}=0$ cases exhibit both significant sensitivity to ${m_l}$ and monotonic dependence, confirming their utility for CRA magnetic moment measurements. 
For $|m_{\gamma}|=\pm 2$ and $\pm 3$ cases, MCVD saturation at 1 makes experimental identification challenging. This behavior contrasts with the case in the first configuration.

In conclusion, the MCVD exhibits a strong dependence on the magnetic moment of Rydberg atoms when interacting with vortex photons carrying specific, low values of TAM. For a general Rydberg atom configuration, the suitable TAM states are $m_{\gamma}=0,\,-1,\,\pm2$. This selection is further restricted to $m_{\gamma}=0,\,-1$ for the CRA configuration.
Experimental investigations should therefore focus primarily on these parameter regimes.

\subsubsection*{phenomenon analyzing}
As established earlier, the $\phi$ integral in Eq.~\ref{vortex} acts as a selection filter through the condition $m_{\gamma}+m_{\Lambda}+m_{\nabla}=0$, where $m_{\Lambda}=0$ or $\pm 1$ originates from the polarization term:
\begin{align}\label{epsilonp}
    \vb{\epsilon}_{\vb{k}{\Lambda_{\gamma}}}\cdot\vb{p} = p_{\perp}\sin^2\frac{\theta_{\vb{k}}}{2}e^{i{\Lambda_{\gamma}}\phi} - p_{\perp}\cos^2\frac{\theta_{\vb{k}}}{2}e^{-i{\Lambda_{\gamma}}\phi} + p_z\sin\theta_{\vb{k}}.
\end{align}
The phase contributions $m_{\nabla}$ emerge from the operator $(-\vb{k}\cdot\vb{\nabla}_{\vb{p}})^{{{\eta}}}$.

{When $|m_{\gamma}|\geq 2$, the dipole contribution (${{\eta}}=0$) vanishes, and the dominant contribution arises from ${{\eta}}=|m_{\nabla}|\geq 1$.
According to Eq. \ref{Rydbergapprox}, four distinct cases emerge:}

{1. When $m_{\gamma}\geq 2$ (implying $m_{\gamma}+m_{\Lambda}>0$ and $m_{\nabla}<0$) and ${m_l}>0$ ($\lambda=1$):
\begin{align}\label{l+}
    (-\vb{k}\cdot\vb{\nabla}_{\vb{p}})^{|m_{\nabla}|} &\rightarrow \left(\frac{k\sin\theta_{\vb{k}}}{p}\right)^{|m_{\nabla}|} e^{-i|m_{\nabla}|\phi}\nonumber\\
    &\quad \times \prod_{j=0}^{|m_{\nabla}|-1}\Big[ \left(l+2+\frac{j}{2}\right) \sin \theta_{\vb p}\Big] 
\end{align}
2. When $m_{\gamma}\geq 2$ (implying $m_{\gamma}+m_{\Lambda}>0$ and $m_{\nabla}<0$) and ${m_l}<0$ ($\lambda=-1$):
\begin{align}\label{l-}
    (-\vb{k}\cdot\vb{\nabla}_{\vb{p}})^{|m_{\nabla}|} &\rightarrow \left(\frac{k\sin\theta_{\vb{k}}}{p}\right)^{|m_{\nabla}|} e^{-i|m_{\nabla}|\phi}\nonumber\\
    &\quad \times \prod_{j=0}^{|m_{\nabla}|-1}\Big[ \left(l+2+\frac{j}{2}\right) \sin \theta_{\vb p}-\frac{l}{\sin\theta_{\vb p}}\Big] 
\end{align}
3. When $m_{\gamma}\leq -2$ (implying $m_{\gamma}+m_{\Lambda}<0$ and $m_{\nabla}>0$) and ${m_l}<0$ ($\lambda=-1$):
\begin{align}\label{l+}
    (-\vb{k}\cdot\vb{\nabla}_{\vb{p}})^{|m_{\nabla}|} &\rightarrow \left(\frac{k\sin\theta_{\vb{k}}}{p}\right)^{|m_{\nabla}|} e^{i|m_{\nabla}|\phi}\nonumber\\
    &\quad \times \prod_{j=0}^{|m_{\nabla}|-1}\Big[ \left(l+2+\frac{j}{2}\right) \sin \theta_{\vb p}\Big] 
\end{align}
4. When $m_{\gamma}\leq -2$ (implying $m_{\gamma}+m_{\Lambda}<0$ and $m_{\nabla}>0$) and ${m_l}>0$ ($\lambda=1$):
\begin{align}\label{l-}
    (-\vb{k}\cdot\vb{\nabla}_{\vb{p}})^{|m_{\nabla}|} &\rightarrow \left(\frac{k\sin\theta_{\vb{k}}}{p}\right)^{|m_{\nabla}|} e^{i|m_{\nabla}|\phi}\nonumber\\
    &\quad \times \prod_{j=0}^{|m_{\nabla}|-1}\Big[ \left(l+2+\frac{j}{2}\right) \sin \theta_{\vb p}-\frac{l}{\sin\theta_{\vb p}}\Big] 
\end{align}
When $m_l$ is close to $l$, the electronic wave function is predominantly localized near $\theta_{\vb{p}} = \pi/2$. In this region, the term $(l + 2 + j/2) \sin \theta_{\vb{p}}$ significantly exceeds $(l + 2 + j/2) \sin \theta_{\vb{p}} - l / \sin \theta_{\vb{p}}$. As a result, when $|m_{\gamma}|$ and $|m_l|$ are fixed, photoionization favors vortex photons that co-rotate with the atomic angular momentum (i.e., $m_{\gamma} > 0$ for $m_l > 0$, and $m_{\gamma} < 0$ for $m_l < 0$). This selectivity leads to MCVD values approaching $\pm 1$, as shown for large $m_l$ in Fig.~\ref{fig:mcd_nl} and for all $m_l$ in Figs.~\ref{fig:mcd_cra} when $|m_{\gamma}|= 2$ and $|m_{\gamma}|=3$. These observations reveal the chiral nature of vortex-photon photoionization from polarized Rydberg atoms.
Conversely, when $m_l$ is much smaller than $l$, the electronic wave function is distributed across the entire angular range $(0, \pi)$. Here, $(l + 2 + j/2) \sin \theta_{\vb{p}}$ is no longer dominant over $(l + 2 + j/2) \sin \theta_{\vb{p}} - l / \sin \theta_{\vb{p}}$, and processes with both co-rotating and counter-rotating vortex photons ($m_{\gamma}$ and $m_l$ having the same or opposite signs) become comparably important. Consequently, the MCVD values lie between $0$ and $\pm 1$, as seen for small $m_l$ with $|m_{\gamma}|= 2$ and $|m_{\gamma}|=3$ in Fig.~\ref{fig:mcd_nl}. }

When $|m_{\gamma}|=1$, the dipole contribution (${{\eta}}=0$) plays important role. Two scenarios occur:

1. When $m_{\gamma}$ and ${\Lambda_{\gamma}}$ share the same sign, the $-p_{\perp}\cos^2(\theta_{\vb{k}}/2)e^{-i{\Lambda_{\gamma}}\phi}$ term dominates, i.e. dipole contribution dominates, yielding MCVD $\approx 0$ resembling initial plane wave case (see $m_{\gamma}=1$ case in Figs.~\ref{fig:mcd_nl} and \ref{fig:mcd_cra}).

 2. When $m_{\gamma}$ and ${\Lambda_{\gamma}}$ have opposite signs, the $p_{\perp}\sin^2(\theta_{\vb{k}}/2)e^{i{\Lambda_{\gamma}}\phi}$ term, which gives dipole contribution, is suppressed by $\sin^2(\theta_{\vb{k}}/2)$. ${{\eta}}=1,2$ contributions (given by $p_z\sin \theta_k$ and $-p_{\perp}\cos^2(\theta_{\vb{k}}/2)e^{-i{\Lambda_{\gamma}}\phi}$ terms respectively) become important. Their interference with dipole contribution introduce significant ${m_l}$-dependent MCVD (see $m_{\gamma}=-1$ case in Figs.~\ref{fig:mcd_nl} and \ref{fig:mcd_cra}).

 When $m_{\gamma}=0$, the dipole contribution is given by $p_z\sin \theta_k$ term and suppressed by $\sin \theta_k$.
 The quadrupole contribution (${{\eta}}=1$) given by $-p_{\perp}\cos^2(\theta_{\vb{k}}/2)e^{-i{\Lambda_{\gamma}}\phi}$ term can be comparable to the dipole contribution. 
 Their interference introduce significant ${m_l}$-dependent MCVD (see $m_{\gamma}=0$ case in Figs.~\ref{fig:mcd_nl} and \ref{fig:mcd_cra}).

\subsection{Photon-energy dependence of MCVD for Rydberg Atoms}
\begin{figure}[htbp]
    \centering
    \includegraphics[width=0.9\linewidth]{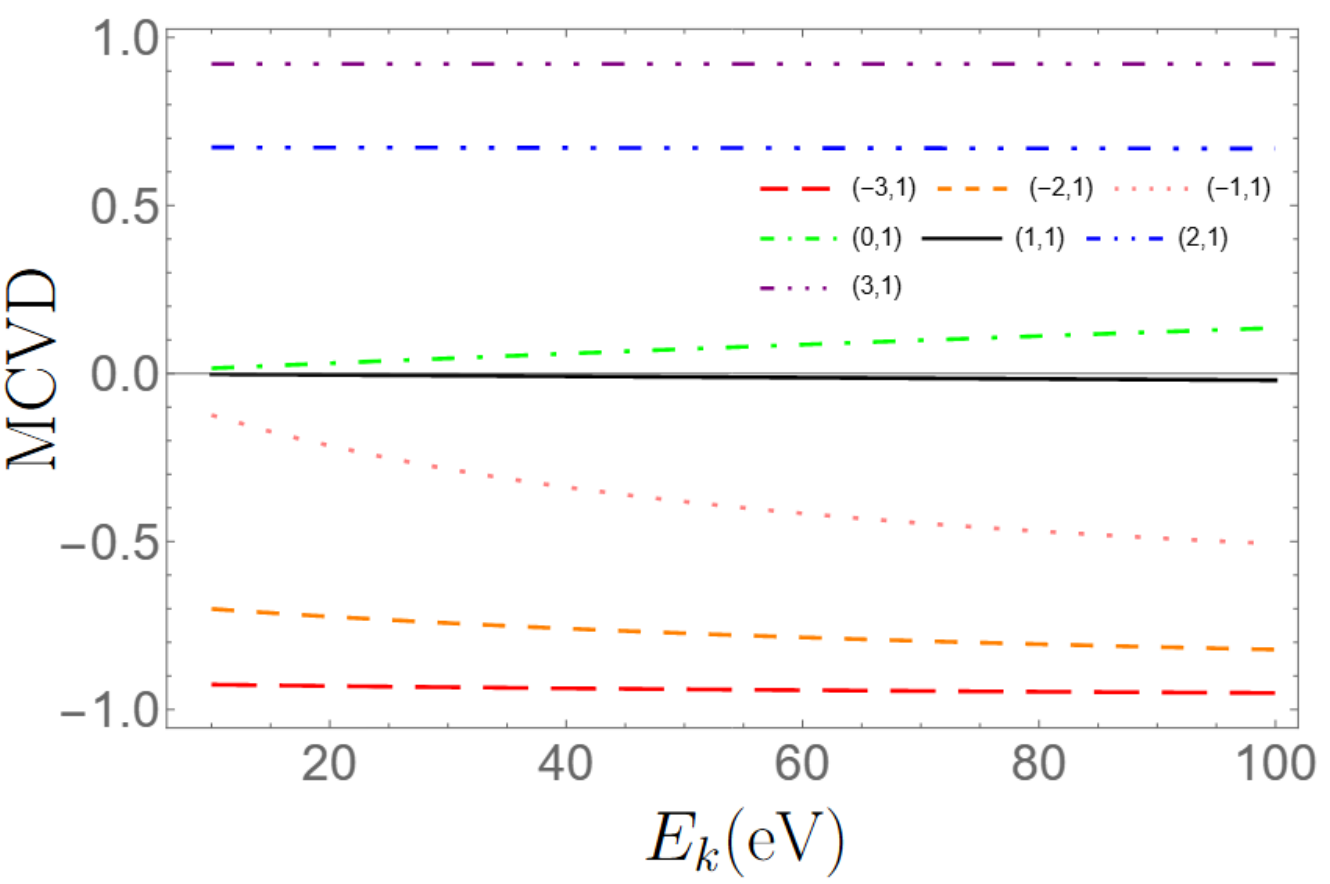}
    \caption{\raggedright MCVD in Rydberg atom photoionization with vortex photons. Targets follow $n=l+1=51$ and ${m_l}=20$. $E_k$ is energy of initial photon ranging from 10 eV to 100 eV.  Conical angle of vortex photon is $\theta_k=\pi/10$. Different lines corresponds to different cases with $m_{\gamma}$ ranging from $0$ to $\pm 3$. Label $(a,b)$ in the legends means we get MCVD from difference of two initial cases: $(m_{\gamma}=a,\Lambda_{\gamma}=b)$ and ($m_{\gamma}=-a,\Lambda_{\gamma}=-b)$.}
    \label{fig:mcvd_energy_n51_m20}
\end{figure}

\begin{figure}[htbp]
    \centering
    \includegraphics[width=0.9\linewidth]{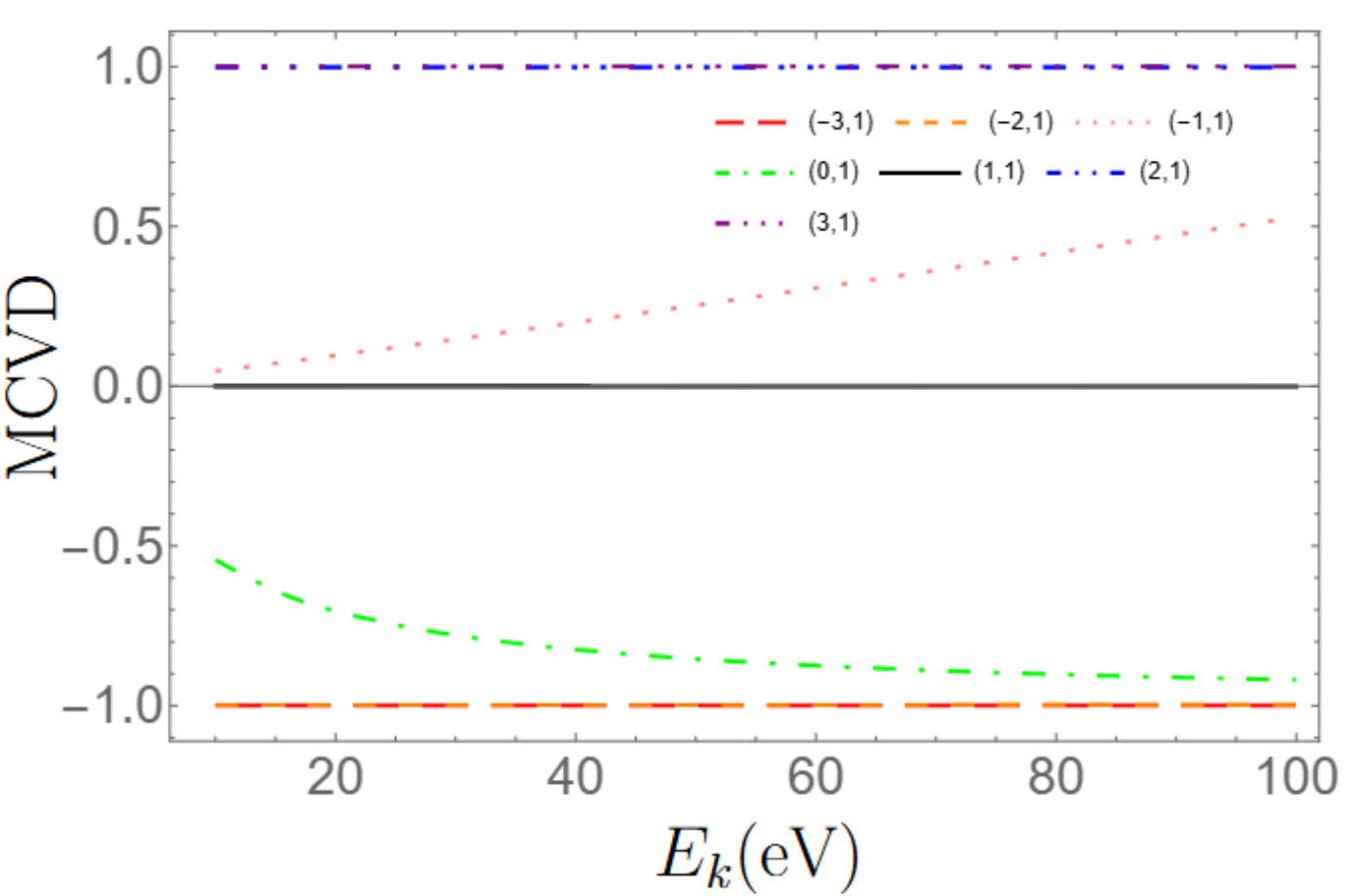}
    \caption{\raggedright MCVD in CRA photoionization with vortex photons. Targets follow $n=l+1={m_l}+1$ with $n=51$. $E_k$ is energy of initial photon ranging from 10 eV to 100 eV. Conical angle of vortex photon is $\theta_k=\pi/10$. Different lines corresponds to different initial energy cases with $m_{\gamma}$ ranging from $0$ to $\pm 3$. Label $(a,b)$ in the legends means we get MCVD from difference of two initial cases: $(m_{\gamma}=a,\Lambda_{\gamma}=b)$ and ($m_{\gamma}=-a,\Lambda_{\gamma}=-b)$.}
    \label{fig:mcvd_energy_n51}
\end{figure}

In addition to the magnetic moment of Rydberg atoms, the MCVD is also influenced by the initial photon energy. Figures~\ref{fig:mcvd_energy_n51_m20} and~\ref{fig:mcvd_energy_n51} demonstrate this dependence for the two configurations defined previously, using vortex photons with TAM quantum numbers $m_\gamma = 0, \pm 1, \pm 2, \pm 3$. 

For the first configuration (Fig.~\ref{fig:mcvd_energy_n51_m20}), with Rydberg atom quantum numbers $n=l+1=51$ and ${m_l}=20$, the MCVD modulus shows distinct behaviors: The cases $m_{\gamma}=-1$ and $m_{\gamma}=0$ exhibit significant enhancement with increasing photon energy, whereas the cases with $m_{\gamma}=1,\,\pm 2,\,\pm 3$ show invisible variation with photon energy varying.

The second configuration (Fig.~\ref{fig:mcvd_energy_n51}), with fixed principal quantum number $n=51$ for CRA, reveals similar trends: The cases $m_{\gamma}=-1$ and $m_{\gamma}=0$ again show strong energy dependence, while the other five cases remain largely unaffected. Notably, cases with $m_{\gamma}=-3$ and $m_{\gamma}=-2$ converge to $-1$ with overlapping curves, and cases with $m_{\gamma}=3$ and $m_{\gamma}=2$ similarly overlap while approaching $1$.

\subsubsection*{phenomenon analyzing}
{These phenomena can be readily understood. As shown earlier, when $|m_{\gamma}| \geq 2$, the dominant {{{{{{multipole}}}}}} contribution has order ${{\eta}} = |m_{\gamma}| - 1$ and scales as $k^{|m_{\gamma}| - 1}$. As a result, the MCVD becomes independent of $k$. Since $E_k = k$ for massless photons, it follows that the MCVD is also independent of the photon energy $E_k$, as illustrated in Fig.~\ref{fig:mcvd_energy_n51_m20} and Fig.~\ref{fig:mcvd_energy_n51} for $|m_{\gamma}| \geq 2$.
For $m_{\gamma} = 0$ or $m_{\gamma} = -1$, both dipole and quadrupole contributions are significant. The dipole term is independent of $k$, while the quadrupole term scales linearly with $k$. This leads to an $E_k$-dependent MCVD, as seen in Fig.~\ref{fig:mcvd_energy_n51_m20} and Fig.~\ref{fig:mcvd_energy_n51} for these cases.
When $m_{\gamma} = 1$, the dominant contribution comes from the dipole term, which is independent of $k$. Consequently, the MCVD remains close to zero, as shown in Fig.~\ref{fig:mcvd_energy_n51_m20} and Fig.~\ref{fig:mcvd_energy_n51} for $m_{\gamma} = 1$.}

These results clearly indicate that the initial photon energy primarily affects MCVD for specific low values of TAM ($m_\gamma = 0,-1$). Since these cases give the most pronounced magnetic moment dependence, careful selection of photon energy becomes crucial. This dual sensitivity underscores the importance of optimal parameter selection for experimental investigations of MCVD.

\subsection{MCVD with different photon TAM $m_{\gamma}$}
\begin{figure}[htbp]
    \centering
        \centering
        \includegraphics[width=0.98\linewidth]{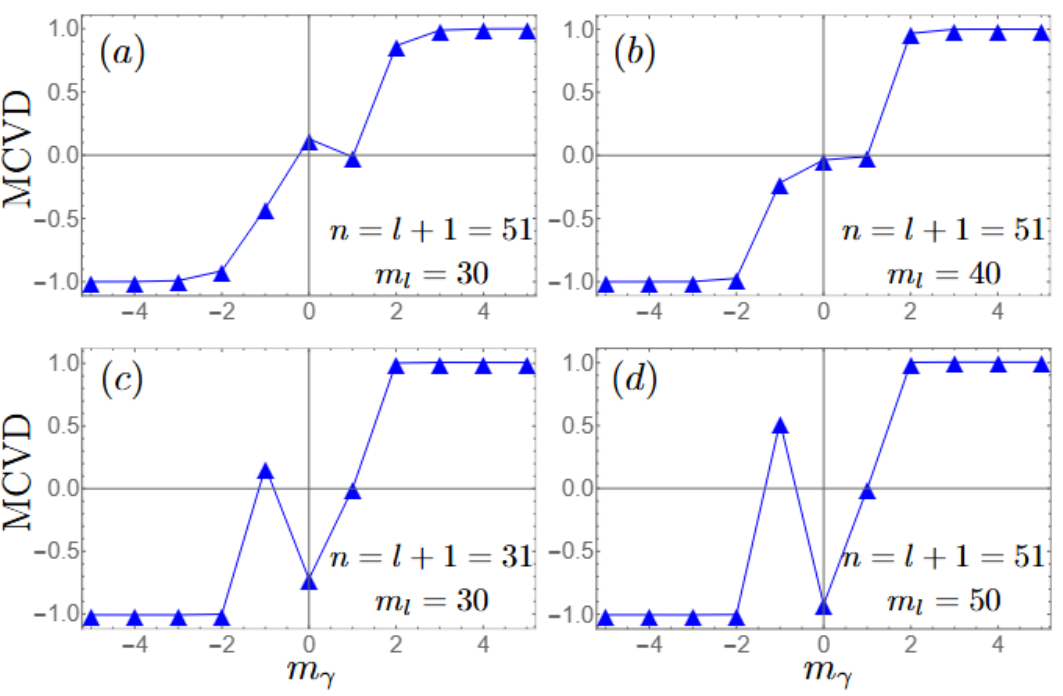}

    \caption{\raggedright Dependence of MCVD on photon TAM quantum number $m_{\gamma}$ for different Rydberg atom configurations: (a) $n=l+1=51$ and $m_l=30$; (b) $n=l+1=51$ and $m_l=40$; (c) CRA with $n=31$; (d) CRA with $n=51$. All cases show $m_{\gamma} \in [-5,5]$. Initial photon energy is chosen to be $100$ eV. Conical angle of vortex photon is $\theta_k=\pi/10$. }
    \label{fig:bigfigure}
\end{figure}

To investigate the dependence of MCVD on the TAM quantum number $m_{\gamma}$ of the initial photon, we analyze fixed Rydberg atom states while varying $m_{\gamma}$. Fig.~\ref{fig:bigfigure} presents four distinct cases:
\begin{itemize}
   \item Subfigures (a) and (b): Rydberg atoms with fixed $n = l + 1 = 51$ at magnetic quantum numbers ${m_l} = 30$ and ${m_l} = 40$ respectively
    \item Subfigures (c) and (d): CRAs (${m_l} = n - 1$) with principal quantum numbers $n = 31$ and $n = 51$ respectively
    
\end{itemize}
All cases examine $m_{\gamma}$ ranging from $-5$ to $5$. Key observations include:

    \begin{itemize}
        \item MCVD $\rightarrow 1$ when $m_{\gamma} \geq 2$
        \item MCVD $\rightarrow -1$ when $m_{\gamma} \leq -2$
        \item MCVD varies between $-1$ and $1$ for $-1 \leq m_{\gamma} \leq 1$
    \end{itemize}

{The asymptotic convergence of MCVD to $\pm 1$ is a characteristic feature for $|m_{\gamma}| \geq 2$. As demonstrated in Figs.~\ref{fig:mcd_nl} and \ref{fig:mcd_cra} for small $|m_{\gamma}|$ values (e.g., 2 and 3), and confirmed in Fig.~\ref{fig:bigfigure} for larger values, this behavior is consistently maintained across all integer values of $|m_{\gamma}| \geq 2$.
Theoretical analysis has been given after Figs.~\ref{fig:mcd_nl} and \ref{fig:mcd_cra}.}
This behavior reveals a fundamental chiral asymmetry in the photoionization dynamics of Rydberg atoms with vortex photons: 
At fixed TAM modulus, for Rydberg states with positive magnetic quantum number (${m_l} > 0$), the ionization cross section is significantly enhanced when interacting with vortex photons carrying positive TAM quantum number ($m_\gamma \geq +2$) compared to those with negative ones ($m_\gamma \leq -2$). 
The parity-conserving interaction rules governing this process show a complementary preference for negative-$m$ states to couple more strongly with negative-$m_\gamma$ photons.

\section{Conclusions}\label{section four}
{This paper examines the photoionization of hydrogen-like Rydberg alkali atoms subjected to vortex photon beams. We propose a novel effect termed magnetic circular vortex dichroism (MCVD) and investigate its behavior across various scenarios. MCVD serves as a key physical observable that provides insights into the characteristics of hydrogen-like Rydberg alkali atoms.}

{
As shown in Figures~\ref{fig:mcd_nl} and~\ref{fig:mcd_cra}, MCVD exhibits high sensitivity to the magnetic moment of Rydberg atoms when the incident vortex photons carry specific low values of total angular momentum (TAM). This sensitivity establishes optical vortices as a powerful tool for probing quantum magnetism in polarized Rydberg systems, paralleling their established role in studying optical magnetism in chiral materials~\cite{sirenko2019terahertz}.
It is important to note that the energy of the incident vortex photon must be carefully chosen to observe significant physical phenomena, as demonstrated in Figures \ref{fig:mcvd_energy_n51_m20} and \ref{fig:mcvd_energy_n51}. Furthermore, we show that the photoionization process exhibits strong angular momentum selectivity, preferentially favoring aligned configurations between photon TAM and atomic magnetic moment. The enhanced cross sections for co-rotating systems versus counter-rotating configurations provide clear evidence of this angular momentum matching. This pronounced asymmetry directly reflects the chiral nature of vortex photon-Rydberg atom collisions.}

{
Experimentally, the MCVD of Rydberg atoms can be measured by irradiating the sample with identical vortex photon beams along the atomic polarization axis from opposite directions. The reversal of the vortex photons' longitudinal momentum does not affect the measurement outcomes.}

\section*{Acknowledgments}
This work was supported in part by Grants No.\,NSFC-12447117. 

\bibliographystyle{utphys} 
\bibliography{refs} 


\begin{appendices}
\section{Calculation of ${\cal M}^V$ with residue theorem}\label{appendix a}
The contour integral in Eq.~\eqref{mvfi}, 
\begin{align}
\oint_{\mathcal{C}} \! dz \, \frac{\mathcal{B}(z)}{(p_{\perp}k_{\perp})^{l+2} z^{\mu-l-m_{\gamma}-{m_l}}(z-z_1)^{l+2}(z-z_2)^{l+2}},
\end{align}
can be evaluated analytically via the residue theorem.

The integral has singularities at:
\begin{itemize}
    \item $z_0 = 0$ (inside unit circle)
    \item $z_1 = \frac{p^2+k^2-2p_zk_z+1/(n^2a^2)-\sqrt{\Delta}}{2p_{\perp}k_{\perp}}$ (inside)
    \item $z_2 = \frac{p^2+k^2-2p_zk_z+1/(n^2a^2)+\sqrt{\Delta}}{2p_{\perp}k_{\perp}}$ (outside)
\end{itemize}
where $\Delta = (p^2+k^2-2p_zk_z+1/(n^2a^2))^2-4p^2_{\perp}k^2_{\perp}$.

For the case that $\Lambda=1$, the transition matrix can be expressed as:
\begin{align}
{\cal M}_{fi}^V &= \frac{i(-1)^n N_BN_P}{2\pi(na)^{2l+4}} e^{i({m_l}+m_{\gamma })\phi_p}\nonumber\\
&\quad \times \sum_{{{\eta}}=0}^l\sum_{{\mu=\mu _{\rm min}}}^{{\mu_{\rm max}}} f_{l{m_l},{{\eta}}\mu}(\vb{p}) (X_a^+ + X_b^+ + X_c^+),
\end{align}
where the residue contributions are:
\begin{align}
X^{+}_a &= \frac{p_{\perp}\sin^2\frac{\theta_{\vb k}}{2}}{(p_{\perp}k_{\perp})^{l+2}} \oint dz \frac{1}{z^{-2-\zeta}(z-z_1)^{l+2}(z-z_2)^{l+2}} \\
X^{+}_b &= \frac{p_{z}\sin\theta_{\vb k}}{(p_{\perp}k_{\perp})^{l+2}}\oint dz \frac{1}{z^{-1-\zeta}(z-z_1)^{l+2}(z-z_2)^{l+2}} \\
X^{+}_c &= \frac{-p_{\perp}\cos^2\frac{\theta_{\vb k}}{2}}{(p_{\perp}k_{\perp})^{l+2}} \oint dz \frac{1}{z^{-\zeta}(z-z_1)^{l+2}(z-z_2)^{l+2}},
\end{align}
where $\zeta ={m_l}+m_{\gamma }+l-\mu$ and $X^+_a$ has residues
\begin{align}
\mathrm{Res}(X^+_a,z_1)
&= \frac{2\pi i p_{\perp} \sin^2\frac{\theta_{\vb k}}{2}\binom{-l-2}{l+1}}{(p_{\perp}k_{\perp})^{l+2}z_1^{-2-\zeta}(z_1-z_2)^{2l+3}} \nonumber \\
& \quad \times {}_2H_1\Big(-2-\zeta,-l-1,-2l-2,1-\frac{z_2}{z_1}\Big) ,
\end{align}
with $_2H_1(a,b,c,d)$ being the Confluent hypergeometric function and
\begin{align}
\mathrm{Res}(X^+_a,z_0)
&=  \frac{2\pi i p_{\perp} \sin^2\frac{\theta_{\vb k}}{2}\binom{-l-2}{-3-\zeta}}{(p_{\perp}k_{\perp})^{l+2}(-z_1)^{l+2}(-z_2)^{\mu-{m_l}+m_{\gamma}-1}} \nonumber \\
&\quad \times {}_2H_1\Big(l+2,3+\zeta,
2+\zeta,\frac{z_2}{z_1}\Big) \nonumber \\
&\text{only for } m_{\gamma} \leq -3-l;
\end{align}

$X^+_b$ has residues
\begin{align}
\mathrm{Res}(X^+_b,z_1)
&=  \frac{2\pi i p_z \sin\theta_{\vb k}\binom{-l-2}{l+1}}{(p_{\perp}k_{\perp})^{l+2}z_1^{-1-\zeta}(z_1-z_2)^{2l+3}} \nonumber \\
&\quad \times {}_2H_1\Big(-1-\zeta,-l-1,-2l-2,1-\frac{z_2}{z_1}\Big)
\end{align}
and
\begin{align}
\mathrm{Res}(X^+_b,z_0)
&=\frac{2\pi i p_z\sin\theta_{\vb k}\binom{-l-2}{-2-\zeta}}{(p_{\perp}k_{\perp})^{l+2}(-z_1)^{l+2}(-z_2)^{\mu-{m_l}+m_{\gamma}}} \nonumber \\
&\quad \times {}_2H_1\Big(l+2,2+\zeta,1+\zeta,\frac{\delta_2}{\delta_1}\Big) \nonumber \\
&\text{only for } m_{\gamma} \leq -2-l;
\end{align}

$X^+_c$ has residues
\begin{align}
\mathrm{Res}(X^+_c,z_1)
&= \frac{-2\pi i p_{\perp} \cos^2\frac{\theta_{\vb k}}{2}\binom{l+1}{-l-2}}{(p_{\perp}k_{\perp})^{l+2}z_1^{-\zeta}(z_1-z_2)^{2l+3}} \nonumber \\
&\quad \times {}_2H_1\Big(-\zeta,-l-1,-2l-2,1-\frac{z_2}{z_1}\Big)
\end{align}
and
\begin{align}
\mathrm{Res}(X^+_c,z_0)
&= \frac{-2\pi i p_{\perp} \cos^2\frac{\theta_{\vb k}}{2}\binom{-l-2}{-1-\zeta}}{(p_{\perp}k_{\perp})^{l+2}(-z_1)^{l+2}(-z_2)^{\mu-{m_l}+m_{\gamma}+1}} \nonumber \\
&\quad \times {}_2H_1\Big(l+2,1+\zeta,\zeta,\frac{z_2}{z_1}\Big) \nonumber \\
&\text{only for } m_{\gamma} \leq -1-l.
\end{align}

For the case that ${\Lambda_{\gamma}}=-1$, we just make exchange  $\sin^2(\theta_{\vb{k}}/2) \rightleftharpoons -\cos^2(\theta_{\vb{k}}/2)$ from ${\Lambda_{\gamma}}=1$ case.

\section{Photon current for vortex photon-Rydberg atom collisions}\label{appendix b}
\subsection{Photon Current in Literature}
Reference \cite{knyazev2018beams} provides detailed expressions for photon current densities in vortex photon collisions with ground state atoms, covering three distinct cases: macroscopic targets, mesoscopic targets, and single-atom targets.

\paragraph{Plane Wave Case}
The photon current density is spatially uniform:
\[ j^{P} = \frac{k}{2\pi}. \]
This expression appears in the denominator of the cross section formula (Eq.~\ref{cross_section}).

\paragraph{Vortex Case -- Macroscopic Target}
Considering Bessel vortex states incident on a macroscopic target (large radius with uniformly distributed atoms), the position-dependent photon current density is:
\begin{align}
j^V(r_{\perp}) &= \frac{k\cos\theta_{\vb k}}{2\pi}\Big[ \cos^4(\theta_{\vb k}/2)J_0^2(k_{\perp}r_{\perp}) \nonumber\\
&\quad + \frac{1}{2}\sin^2\theta_{\vb k} J_1^2(k_{\perp}r_{\perp}) + \sin^4(\theta_{\vb k}/2)J^2_2(k_{\perp}r_{\perp})\Big],\
\end{align}
where $r_{\perp}$ denotes the transverse coordinate. In the limit $\theta_{\vb k}\rightarrow 0$, $j^V(r_{\perp})\rightarrow j^{P}$.

For cross section calculations, we use the coordinate-independent effective current:
\[ j^V_{\text{mac}} = \int j^V(b)d^2\vec{b} = \frac{Rk\cos\theta_{\vb k}}{\pi k_{\perp}}, \]
where $R$ is the target radius and $\vec{b}$ represents the impact parameter. The radius $R$ becomes irrelevant for sufficiently large targets.

\paragraph{Vortex Case -- Mesoscopic and Single-Atom Targets}
For mesoscopic targets (small radius with atomic distribution $n(\vec{r}_{\perp})$), the effective photon current becomes:
\[ j^V_{\text{mes}} = \int j^V(r_{\perp})n(\vec{r}_{\perp})d^2\vec{r}_{\perp}. \]

For a single-atom target treated as a point particle (e.g., a ground-state atom), we take $n(\vec{r}_{\perp}) = \delta(r_{\perp})$, yielding:
\[ j^V_{\text{sin}} = \int j^V(r_{\perp})\delta(r_{\perp})d^2\vec{r}_{\perp} = \frac{k\cos\theta_{\vb k}\cos^4(\theta_{\vb k}/2)}{2\pi}. \]

\subsection{Photon Current for Vortex Beams with Rydberg Atom Targets}
Rydberg atoms, having large spatial extent, cannot be treated as point particles. Following the methodology from previous cases, we define the effective photon current as:
\[ j^V_{R} = \int j^V(r_{\perp})|\psi_i|^2(\vec{r})d^3\vb{r}. \]

This formulation correctly reduces to:
\begin{itemize}
    \item Plane wave case (through wavefunction normalization)
    \item Macroscopic case: $j^V_{\text{mac}} = \int j^V(r_{\perp})\delta(z)d^3\vb{r}$
    \item Mesoscopic case: $j^V_{\text{mes}} = \int j^V(r_{\perp})n(\vec{r}_{\perp})\delta(z)d^3\vb{r}$
\end{itemize}
The $\delta(z)$ terms reflect the longitudinal localization of conventional atomic targets. Physically, the effective current represents the photon flux experienced by the target.
\end{appendices}

\end{document}